\documentclass[12pt]{article}
\usepackage{graphicx}

\makeatletter
\newcommand{\fnsy}{}

\renewcommand{\@makefnmark}{\hbox{\mathsurround=0pt $^{\fnsy}$}}
\renewcommand{\@makefntext}[1]{\parindent=1em\noindent\hbox to 1.8em
    {\hss$^{\fnsy}$}#1}

\makeatother

\font\tit=cmss17 scaled \magstep5
\font\nam=cmssbx10 scaled \magstep4
\font\adsl=cmssi10 scaled \magstep4
\font\nida=cmss10 scaled \magstep4
\font\nidas=cmss10 scaled \magstep2

\font\nidabox=cmssq8 scaled \magstep5
\font\nidaleg=cmss10 scaled \magstep3

\begin{document}

\baselineskip=1.5\baselineskip

\parindent=0mm

\renewcommand{\fnsy}{{\LARGE *}}

\begin{center}
{\tit
Information   processing at\\  single neuron level$^{\makebox{\tit *}}$
\footnotetext[1]{Poster presented at NATO ASI
Modulation of Neuronal Signaling: Implications for Visual Perception,
Nida, Lithuania, 12-21 July, 2000. See also
Vidybida A.K. Inhibition as binding controller at the single
neuron level. BioSystems, Vol. 48, 1998, p. 263--267;
Vidybida A.K. Neuron as time coherence discriminator. Biological
Cybernetics, 74(6), 1996, 539-544.
}} \end{center} \bigskip\medskip\vspace{5mm}

{\nam A.K.Vidybida}\medskip

{\adsl Bogolyubov Institute for Theoretical Physics\\
03680 Kyiv, Ukraine\\
E-mail: vidybida@bitp.kiev.ua\\
http://www.bitp.kiev.ua/pers/vidybida}
\bigskip\medskip\vspace{10mm}

\nida
\hspace*{\baselineskip}
The understanding of mechanisms of higher brain functions expects a continuous
reduction from higher activities to lower ones, eventually, to activities in
individual neurons, expressed in terms of membrane potentials and ionic
currents. While this approach is correct scientifically and desirable for
applications, the complete range of the reduction is unavailable to a single
researcher due to human brain limited capacity. In this connection, it would be
helpful to abstract from the rules by which a neuron changes its membrane
potentials to rules by which the information is processed in the neuron. The
``coincidence detector", and ``temporal integrator" are the examples of such an
abstraction.

\newpage
\baselineskip=0.667\baselineskip

\nida
\baselineskip=1.64\baselineskip
\parskip=0.1\baselineskip

While being useful in constructing artificial networks, the above two
abstractions are neither connected with known brain functions, nor they are
relevant to the biological nature of nervous cell, including
justification of its survival in
the natural selection.

In this poster, an alternative abstraction is described, which seems to be free
from these shortages. Based on the Hodgkin and Huxley set of equations, we
analyze the neuronal reaction to  compound stimuli, comprising large number
(1000) of EPSP (Box 1). The unitary EPSPs in a compound stimulus are
distributed randomly over a time window [0;W]. The probability to fire a spike
as a function of W is calculated by means of Monte Carlo method. In this
course, several values of proximal GABA$_{\makebox{\nidas b}}$-type inhibition
are applied (Box 1, term with {\LARGE $g_{iK}$}).
  The dependencies obtained (Fig.4) allow to formulate one
more abstraction of neuronal functioning (Box 2). In this abstraction the
temporal structure of stimulus as well as the inhibition get their information
processing meaning.  The formulation is expressed in terms of binding, or
feature linking --- an essential ability of the brain. In this formulation a
single neuron is endowed with a meaningful ability, the binding, which might be
the reason for survival of excitable cells in the natural selection. The
information processing scheme in a single neuron (Fig.5) could be the first
step in the bottom-up reductionistic explanation of brain functioning.

\newpage
\nida

\vspace*{-2.0\baselineskip}
\hspace*{0.0\baselineskip}
\input{fig1.pic}

\noindent Fig.1. Time course of (1) excitatory sinaptic current in the sinaptic
part of membrane, ESC, and (2) excitatory postsynaptic potential in the
triggering zone, EPSP
\vspace*{2\baselineskip}

\includegraphics[width=0.8\textwidth]{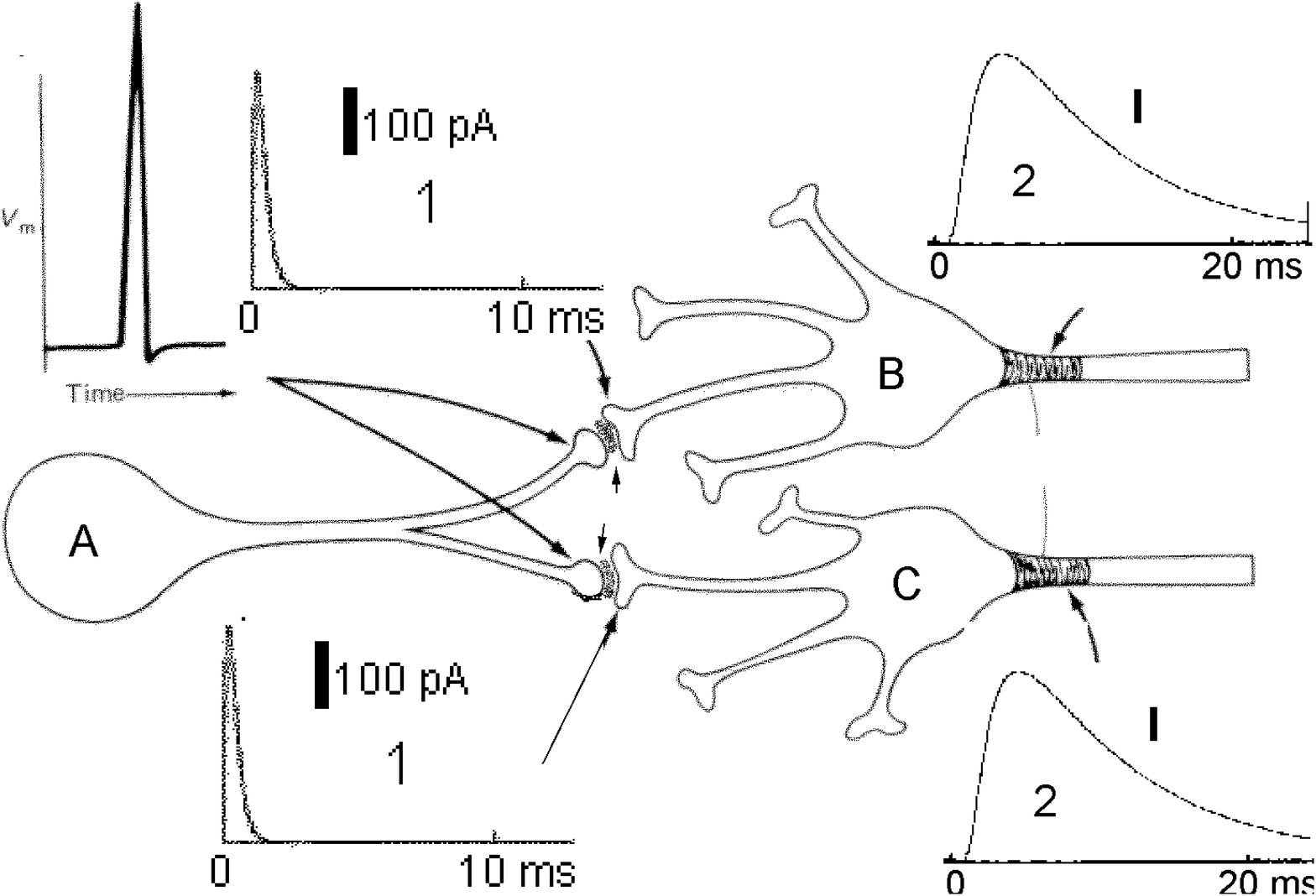}
\unitlength=1mm
\begin{picture}(0,0)(139,-1.5)
\put(33,80){\makebox{\small ESC\hfill\\}}
\put(30,27){\makebox{\small ESC\hfill\\}}
\put(20,90){\makebox{\small Spike\hfill\\}}
\put(63,34){\makebox{\small Synapses\hfill\\}}
\put(105,40){\makebox{\small triggering zone\hfill\\}}
\put(122,14){\makebox{\small 10 $\mu$V\hfill\\}}
\put(122,74){\makebox{\small 10 $\mu$V\hfill\\}}
\put(100,83){\makebox{\small EPSP\hfill\\}}
\put(103.5,21){\makebox{\small EPSP\hfill\\}}
\end{picture}
\medskip

\noindent Fig.2. Mechanism of the EPSP forming {\large
(modified from:
Principles of Neural
Science, E. Kandel and J. Schwartz(eds), Elsevier, 1985)}

\newpage

\centerline{\nidabox Box 1}
\bigskip\medskip

\hspace*{-1.5\baselineskip}
\nida
\fboxrule=1pt
\framebox{
\parbox{\textwidth}{
\begin{center}
\nida Hodgkin and Huxley equations
\end{center}

\LARGE
$$\begin{array}{c}
C_MdV/dt=\\
-g_{K}n^{4}(V-V_{K}) -
 g_{Na}m^{3}h(V-V_{Na}) - \\
-g_{l}(V-V_{l})-g_{iK}(V-V_{K})+\\
+I(t),
\end{array}$$

$$dn/dt=\alpha_{n}(1-n)-\beta_{n}n ,
  $$$$
dm/dt=\alpha_{m}(1-m)-\beta_{m}m ,
      $$
$$dh/dt=\alpha_{h}(1-h)-\beta_{h}h .
  $$
 \bigskip

$$
I(t)=-C_{M}dCompEPSP(t)/dt.
$$
\centerline{\nidaleg Stimulating current in the triggering zone,}
\centerline{\nidaleg which is used for numerical simulation}
 \bigskip

$$
     CompEPSP(t)=\sum_{k=1}^{1000}EPSP(t-t_{k})
$$
\begin{center}\nidaleg Compound potential in the triggering zone.
All $\displaystyle t_k$ are chosen randomly from the time window:
$\displaystyle t_k\in [0;W]$. See example in Fig.3.
 \underline{Temporal coherence},  $\displaystyle
TC$, is defined as $\displaystyle TC= 1/W$.  \end{center} \bigskip

                   }
         }

\newpage

\nida

\vspace*{-2.0\baselineskip}
\hspace*{0.0\baselineskip}
\input{fig3.pic}

\noindent Fig.3. Example of compound stimulus (solid line)
comprising three unitary stimuli (dotted lines) as they are seen
in the triggering zone
\vspace*{1\baselineskip}

\input{fig4.pic}

\noindent Fig.4.
Firing probability vs temporal coherence between the unitary stimuli
within the compound stimulus comprising 1000 of unitary stimuli. The four curves
correspond consecutively from the left to the right to the inhibition
potentials 0.43, 3.08, 5.02, 6.30 \LARGE$mV$.

\newpage

\vspace*{-3\baselineskip}
\centerline{\nidabox Box 2}
\bigskip\medskip

\hspace*{-1.5\baselineskip}
\nida
\fboxrule=1pt
\framebox{
\parbox{\textwidth}{
\begin{center}
Information processing in a generic neuron
\end{center}

\begin{enumerate}
\item Excitatory synaptic currents (ESCs, Fig.1) are trea\-ted as elementary
events registered by the neuron.\phantom{d}
\item EPSP, which follows the ESC serves as short term memory
mechanism, because its duration is much longer than that of ESC (Fig.1).
\item A set of elementary events, which are
coherent in time, is bound in the
neuron into an output spike, which represents the bound event (Fig.5).
\item Inhibition serves as controller of this type of binding
(Figs.4,5).
\end{enumerate}
                   }
         }
\vspace{1mm}

\large
\hspace*{\baselineskip}
\input{scheme.pic}
\vspace{4mm}

\nida\noindent
Fig.5. Proposed scheme of information processing in a single generic neuron

\end{document}